# Robust topological state against magnetic impurities observed in superconductor PbTaSe$_2$

**Authors**
Daniel Multer[1], Jia-Xin Yin[1†], Songtian S. Zhang[1], Hao Zheng[2], Tay-Rong Chang[3,4,5], Guang Bian[6], Raman Sankar[7,8], M. Zahid Hasan[1,9*]

**Affiliations**
[1]Laboratory for Topological Quantum Matter and Advanced Spectroscopy, Department of Physics, Princeton University, Princeton, New Jersey 08544, USA.

[2]Department of Physics and Astronomy, Shanghai Jiaotong University, Shanghai 200240, China.

[3]Department of Physics, National Cheng Kung University, Tainan 701, Taiwan.

[4]Center for Quantum Frontiers of Research & Technology (QFort), Tainan 701, Taiwan

[5]Physics Division, National Center for Theoretical Sciences, National Taiwan University, Taipei, , Taiwan.

[6]Department of Physics and Astronomy, University of Missouri, Columbia, Missouri 65211, USA.

[7]Institute of Physics, Academia Sinica, Taipei 11529, Taiwan.

[8]Center for Condensed Matter Science, National Taiwan University, Taipei 10617, Taiwan.

[9]Lawrence Berkeley National Laboratory, Berkeley, California 94720, USA.

†Corresponding authors. e-mail: jiaxiny@princeton.edu; mzhasan@princeton.edu

**Magnetic impurities deposited on topological superconductor candidate PbTaSe$_2$ can introduce a non-splitting zero-energy state inside the superconducting gap, which has been proposed as a field-free platform for topological zero modes. However, it is still unclear how robust the topological state in PbTaSe$_2$ is against magnetic impurities, which is related to the topological nature of the zero-energy state as well as its potential for quantum computation. In this work, we use scanning tunneling microscopy (STM) to study the topological surface state in the normal state of PbTaSe$_2$ under the perturbation of magnetic impurities. We visualize the quasi-particle interference (QPI) arising from the topological surface state. We then deposit Fe impurities on the surface to form atomic Fe adatoms. We find that each Fe adatom sits at a unique interstitial position on the surface and features a local state at high energies, both of which are consistent with our first-principles calculation that further reveals its large magnetic moment. Our systematic Fe deposition and subsequent measurements show that the arc-like QPI pattern at the Fermi energy is robust with up to 3% Fe coverage where the atomic nature of Fe adatoms still holds. Our results provide evidence that topological surface state at the Fermi energy in PbTaSe$_2$ is robust against dilute magnetic impurities.**

The interplay between magnetism, topology, and superconductivity is at the quantum frontier with many open questions [1-4], where inducing magnetism in a superconductor with a topological band structure is a promising research direction [5]. PbTaSe$_2$ is a non-centrosymmetric superconductor [6-12] with the space group of $P\bar{6}m2$ (No.187) and superconducting transition temperature of 3.8K [Fig. 1(a)]. Theoretical calculations indicate a non-zero $Z_2$ topological invariant in the band structure [9], featuring a spin-helical surface state near the Fermi level, as shown in Fig. 1(b). We observe this pronounced surface state in angle-resolved photoemission consistent with previous measurements [8], and STM measurements have also demonstrated such topological surface states with fully gapped bulk superconductivity [10]. Moreover, deep in the superconducting state, STM has shown a robust zero-energy peak in both the magnetic field indued vortex core [10] and Fe adatom through atomic deposition [12], making it a tantalizing topological superconductor platform for visualizing Majorana zero modes. However, while the production of Majorana zero modes requires the coexistence of magnetism, topological surface state, and superconductivity, it is largely unclear how robust the topological surface state against magnetic perturbations. In this work, we provide experimental evidence to show that the topological surface state is robust against deposited atomic magnetic impurities up to 3% surface concentrations.

$PbTaSe_2$ has a layered crystal structure consisting of hexagonal lattices [Fig. 1(c)]. Cryogenic cleaving produces either a Pb surface or a Se surface [10]. As the Se surface is found to have certain reconstructions, and we, therefore, focus on the unreconstructed Pb surface in this work [Fig. 1(d)]. Single crystals of $PbTaSe_2$ were cleaved at 77K in ultra-high vacuum. Our STM measurements were all performed at 4.2K. Topographic images were taken with bias voltage V=-100mV and tunneling current I=0.1nA. Differential conductance dI/dV maps for different Fe concentrations were taken at V=-100mV and I=0.3nA with bias modulation of 5mV. dI/dV map for imaging the wider dispersion of the pristine sample was taken at V=-1V and I=1nA with bias modulation of 15mV. Magnetic fields were applied under zero-field cooling. We deposit Fe (99.999% purity) while holding the temperature of the cleaved crystal at 77K. By controlling the deposition rate, we can obtain various adatom Fe surface concentrations, ranging from 0 to 3%, where the atomic nature of the deposited Fe still holds (no clusters). We heat a Fe wire (diameter of 0.5mm) by applying a direct current of 2A as the deposition source, which is located 10cm away from the surface of the sample cooled at 77K. The deposition rate is found to be around 2% surface concentration per minute, and we control the surface concentration of Fe by controlling the deposition time. We estimate the concentration of Fe adatoms by counting the numbers of Fe adatoms relative to the lattice atoms for an area of 200Å×200Å. The error bar of the concentration is estimated to be around ±0.2%. In order to determine the exact location and configuration of the Fe atoms, we compute electronic structures using the projector augmented wave method [13,14] as implemented in the VASP package [15,16,17] within the generalized gradient approximation schemes [18]. The in-plane 4×4 supercell lattice structures with 4-unit-cell symmetric slabs were used to simulate the single Fe impurity effect. A 5×5×1 Monkhorst Pack k-point mesh was used in the computations. The spin-orbit coupling effects were included self-consistently.

We find that the native Pb vacancies on the Pb surface can induce standing waves [Fig. 1(e)]. The standing waves propagate anisotropically instead of simple ring-like signals, indicative of nontrivial scattering processes [4]. We further perform dI/dV mapping on a large Pb surface with native Pb vacancies as shown in Figs. 1(f) and (g). By performing a Fourier transform of the dI/dV map data, we obtain the QPI signal. The QPI data at Fermi energy manifests itself as disconnected arcs with six-fold symmetry [Fig. 1(g) inset]. This distinct feature has been identified as a fingerprint for the topological surface states in this material as well as others, including topological insulators [4,18-25]. It arises from the warping of a closed Fermi surface with helical spin texture [26], where the quasi-particle scattering is allowed only along the arc directions. Compared with previous QPI studies [10], we do not observe a clear splitting of the arc feature. The QPI dispersion shown in Fig. 1(h) also confirms that it arises from a Dirac-like band with the Dirac cone energy around 1eV, consistent with the first-principles calculation [9,10].

Our previous experiments have shown that deep in the superconducting state, each Fe adatom locally induces a zero-energy bound state, with each Fe adatom being a strong magnetic impurity [12]. We further investigate its magnetic nature and its impact on the topological surface state. Figure 2(a) shows a typical atomically resolved topographic image of the deposited Fe atoms on Pb surface, with 0.5% coverage. The deposited Fe atoms form randomly scattered atomic adatom impurities, which appear in the STM scan as bright spots. Each Fe adatom sits in the interstitial position around the center of three Pb surface atoms [Fig. 2(b)]. Moreover, owing to the non-centrosymmetry of the crystal, there should be two inequivalent interstitial positions, which can be noted as α and β. One should be just above the underlying Ta atoms, while the other is not. Analysis of the atomic positions of the Fe adatoms in Fig. 2(c) reveals they are all located in the same position, which we note as α. The fact that all Fe adatoms sit at α suggests that this configuration has the lowest energy, although the experimental data by itself does not determine whether or not that site is the one above the underlying Ta atoms. Moreover, through the wide-energy dI/dV spectroscopy, we find each Fe adatom induces additional states near -2.5eV [Fig. 2(d)]. The atomic position and induced states of Fe adatom observed here offer constraints for the first-principles study of its magnetism.

By calculating the energy configurations and magnetic moments for different positions as illustrated in the lower panel of Fig. 2(e), we find that when the Fe atom is at the interstitial position that is above a Ta atom, the total energy is substantially minimized [Fig. 2(e)]. This calculation results support that the experimentally observed α position to be that above a Ta atom. Moreover, at this position, Fe carries a large magnetic moment of $4.7\mu_B$, much higher than the moment of Fe in bulk as $2.2\mu_B$. To understand this unusually strong magnetism, we calculate the spin and orbital resolved density of states for Fe atom at this position, the Pb atom nearby, and the Pb atom far away from Fe, respectively, as shown in Fig. 2(f). Consistent with our experiments, the calculated density of states near the Fe impurity has additional states near -2.5eV. In our calculation for states of Fe and far away Pb atoms, there are also less pronounced differences at other energies, which are not clearly detected in our experiments. We further find that the charge number of the Pb atom (calculated to be 3.2 based on the integration of occupied density of states) near Fe is larger than other Pb atoms (calculated to be 2.9) on the surface. In addition, the $p$ orbitals of nearby Pb atoms share similar energy levels as the $d$ orbitals of Fe at -2.5eV. These findings suggest that the electrons transfer from Fe to Pb due to the $d$-$p$ hybridization, causing large exchange splitting in Fe $d$-orbital, and consequently a large magnetic moment. In particular, the vertical $d$ orbital can hybridize strongly with planar $p$ orbitals, and the splitting of $d_{z2}$ orbital is larger than others. We also evaluate the spin anisotropy energy by calculating the total energy of the system with aligning the spin of Fe adatom either along the $c$-axis ($S_Z$) or the $a$-axis ($S_X$). We find that the energy of spin along $S_Z$ is slightly lower than $S_X$, which is around 7meV, indicating the spin favors the z-direction.

Having determined the atomic and magnetic characteristics of the individual impurity, we now examine the impact of magnetic impurities on the topological surface state. Taking differential conductance maps for each impurity concentration [Fig. 3(a)], we analyze the QPI patterns derived from the fast Fourier transforms. We find that while the arc-like QPI signal broadens with increasing Fe adatom concentration, no obvious new scattering vectors [Fig. 3(b)] or changes in the dispersion pocket [Fig. 3(c)] are introduced by the atomic Fe impurities. These observations suggest that despite some decoherence, the topological surface states are robust against atomic Fe impurities up to 3% and that the spin-momentum locked electronic structure around $E_F$ remains unaltered. This observation of robust arc-like QPI pattern is similar to that observed in magnetic topological insulators $Bi_{2-x}Mn_xTe_3$ [27], but different from that observed in $Bi_{2-x}Fe_xTe_3$ and $Bi_2Te_3$ with surface Mn impurities where new scattering channels are created by magnetic impurities [28,29]. As can be seen in our data, the surface state dispersion is very steep with a large Fermi velocity of 2.5eV•Å, and the Dirac cone is located far from the Fermi level (about +1eV). The magnetism that breaks time-reversal symmetry can strongly affect the spin-orbit structure near the surface Dirac cone but only weakly perturb the state far from it, which is consistent with our observation that the topological surface state near the Fermi level is robust. We also note that our QPI dispersion data in Fig. 3(c) reveals a broad cone-like feature, with the cone energy being around 65meV for all concentrations. The origin of this feature is unclear to us, and whether it arises from the surface projection of the bulk Dirac-like bands (for instance, the Dirac nodal line) deserves further attention.

The experiments demonstrated here suggest that the low-energy topological surface state with a helical spin texture can be taken as an invariant in studying this superconductor with dilute magnetic impurities. This observation is consistent with the topological interpretation of the non-splitting zero-energy states observed at Fe adatoms in several topological superconductor candidates, including Fe(Te,Se), LiFeAs, and $PbTaSe_2$ [12, 30, 31, 32, 33], and encourages further exploration of topological zero modes in similar platforms with a combination of topology, magnetism, and quantum ordering [34, 35, 36].

**References:**

[1] S. R. Elliott and M. Franz, Rev. Mod. Phys. 87, 137 (2015).

[2] C. W. J. Beenakker, Annu. Rev. Condens. Matter Phys. 4, 113 (2013).

[3] M. Sato and Y. Ando, Reports Prog. Phys. 80, 076501 (2017).

[4] J-X. Yin, S.H. Pan and M Zahid Hasan, Nature Review Physics 3, 249 (2021).

[5] L. Fu and C. L. Kane, Phys. Rev. Lett. 100, 096407 (2008).

[6] R. Eppinga and G. Wiegers, Physica B+C (Amsterdam) 99,121(1980).

[7] M. N. Ali, Q. D. Gibson, T. Klimczuk and R. J. Cava, Phys. Rev. B. 89, 020505 (2014).

[8] G. Bian, T.-R. Chang, R. Sankar, S.-Y. Xu, H. Zheng, T. Neupert, C.-K. Chiu, S.-M. Huang, G. Chang, I. Belopolski *et al*., Nat. Commun. 7, 10556 (2016).

[9] T.-R. Chang, P.-J. Chen, G. Bian, S.-M. Huang, H. Zheng, T. Neupert, R. Sankar, S.-Y. Xu, I. Belopolski, G. Chang *et al*., Phys. Rev. B 93, 245130 (2016).

[10] S.-Y. Guan, P.-J. Chen, M.-W. Chu, R. Sankar, F. Chou, H.-T. Jeng, C.-S. Chang and T.-M. Chuang, Sci. Adv. 2, e1600894–e1600894 (2016).

[11] C.-L. Zhang, Z. Yuan, G. Bian, S.-Y. Xu, X. Zhang, M. Z. Hasan and S. Jia, Phys. Rev. B. 93, 054520 (2016).

[12] S. S. Zhang, J.-X. Yin, G. Dai, L. Zhao, T.-R. Chang, N. Shumiya, K. Jiang, H. Zheng, G. Bian, D. Multer, *et al*., Phys. Rev. B 101, 100507(R) (2020).

[13] P. E. Blöchl, Phys. Rev. B. 50, 17953 (1994).

[14] G. Kresse, D. Joubert, Phys. Rev. B. 59, 1758 (1999).

[15] G. Kresse, J. Hafner, Ab initio molecular dynamics for open-shell transition metals. Phys. Rev. B. 48, 13115 (1993).

[16] G. Kresse, J. Furthmüller, Comput. Mater. Sci. 6, 15 (1996).

[17] G. Kresse, J. Furthmüller, Phys. Rev. B. 54, 11169 (1996).

[18] J. P. Perdew, K. Burke, M. Ernzerhof, Phys. Rev. Lett. 77, 3865 (1996).

[19] P. Roushan, J. Seo, C. V. Parker, Y. S. Hor, D. Hsieh, D. Qian, A. Richardella, M. Z. Hasan, R. J. Cava, and A. Yasdani, Nature 460, 1106–1109 (2009).

[20] T. Zhang, P. Cheng, X. Chen, J.-F. Jia, X. Ma, K. He, L. Wang, H. Zhang, X. Dai, Z. Fang, *et al*., Phys. Rev. Lett. 103, 266803 (2009).

[21] H. Zheng, S.-Y. Xu, G. Bian, C. Guo, G. Chang, D. S. Sanchez, I. Belopolski, C.-C. Lee, S. M. Huang, X. Zhang, *et al*., ACS Nano 10, 1378 (2016).

[22] H. Zheng et al., Phys. Rev. Lett. 117, 266804 (2016).

[23] H. Zheng, G. Bian, G. Chang, H. Lu, S.-Y. Xu, G. Wang, T.-R. Chang, S. Zhang, I. Belopolski, N. Alidoust, *et al*., Phys. Rev. Lett. 117, 266804 (2016).

[24] J.-X. Yin, S. S. Zhang, H. Li, K. Jiang, G. Chang, B. Zhang, B. Lian, C. Xiang, I. Belopolski, H. Zheng, *et al*., Nature 562, 91-95 (2018).

[25] J.-X. Yin, Nana Shumiya, Sougata Mardanya, Qi Wang, Songtian S. Zhang, Hung-Ju Tien, Daniel Multer, Yuxiao Jiang, Guangming Cheng, Nan Yao et al., Nature Communications 11, 4003 (2020).

[26] L. Fu, Phys. Rev. Lett. 103, 1 (2009).

[27] H. Beidenkopf, P. Roushan, J. Seo, L. Gorman, I. Drozdov, Y. S. Hor, R. J. Cava, and A. Yasdani, Nat. Phys. 7, 939 (2011).

[28] Y. Okada, C. Dhital, W. Zhou, E. D. Huemiller, H. Lin, S. Basak, A. Bansil, Y.-B. Huang, H. Ding, Z. Wang et al., Phys. Rev. Lett. 106, 206805 (2011).

[29] P. Sessi, F. Reis, T. Bathon, K. A. Kokh, O. E. Tereshchenko and M. Bode, Nat. Commun. 5, 5349 (2014).

[30] J.-X. Yin, Z. Wu, J.-H. Wang, Z.-Y. Ye, J. Gong, X.-Y. Hou, L. Shan, A. Li, X.-J. Liang, X.-X. Wu, et al., Nat. Phys. 11, 543 (2015).

[31] Jia-Xin Yin, Songtian S. Zhang, Guangyang Dai, Yuanyuan Zhao, Andreas Kreisel, Gennevieve Macam, Xianxin Wu, Hu Miao, Zhi-Quan Huang, Johannes H. J. Martiny et al., Phys. Rev. Lett. 123, 217004 (2019).

[32] Chaofei Liu, C. Chen, X. Liu, Z. Wang, Y. Liu, S. Ye, Z. Wang, J. Hu, and J. Wang, Sci. Adv. 6, eaax7547 (2020).

[33] P. Fan, F. Yang, G. Qian, H. Chen, Y.-Y. Zhang, G. Li, Z. Huang, Y. Xing, L. Kong, W. Liu, *et al*., Nat. Commun. 12, 1348 (2021).

[34] J.-X. Yin, Songtian S. Zhang, Guoqing Chang, Qi Wang, Stepan Tsirkin, Zurab Guguchia, Biao Lian, Huibin Zhou, Kun Jiang, Ilya Belopolski et al., Nat. Phys. 15, 443-448 (2019).

[35] J.-X. Yin, Wenlong Ma, Tyler A. Cochran, Xitong Xu, Songtian S. Zhang, Hung-Ju Tien, Nana Shumiya, Guangming Cheng, Kun Jiang, Biao Lian et al., Nature 583, 533 (2020).

[36] Yu-Xiao Jiang, Jia-Xin Yin, M. Michael Denner, Nana Shumiya, Brenden R. Ortiz, Gang Xu, Zurab Guguchia, Junyi He, Md Shafayat Hossain et al., Nat. Mater. (2021). https://doi.org/10.1038/s41563-021-01034-y

**Acknowledgement:** Work at Princeton University was supported by the Gordon and Betty Moore Foundation (GBMF4547 and GBMF9461; M.Z.H.). The theoretical work and sample characterization are supported by the United States Department of Energy (U.S. DOE) under the Basic Energy Sciences programme (grant number DOE/BES DE- FG-02-05ER46200; M.Z.H.). The work on topological superconductivity is partly based on support by the U.S. DOE, Office of Science through the Quantum Science Center (QSC), a National Quantum Information Science Research Center at the Oak Ridge National Laboratory. T.-R.C. was supported by the Young Scholar Fellowship Program from the Ministry of Science and Technology (MOST) in Taiwan, under a MOST grant for the Columbus Program MOST110-2636-M-006-016, NCKU, Taiwan, and National Center for Theoretical Sciences, Taiwan. Work at NCKU was supported by the MOST, Taiwan, under grant MOST107-2627-E-006-001 and Higher Education Sprout Project, Ministry of Education to the Headquarters of University Advancement at NCKU.

**Figures:**

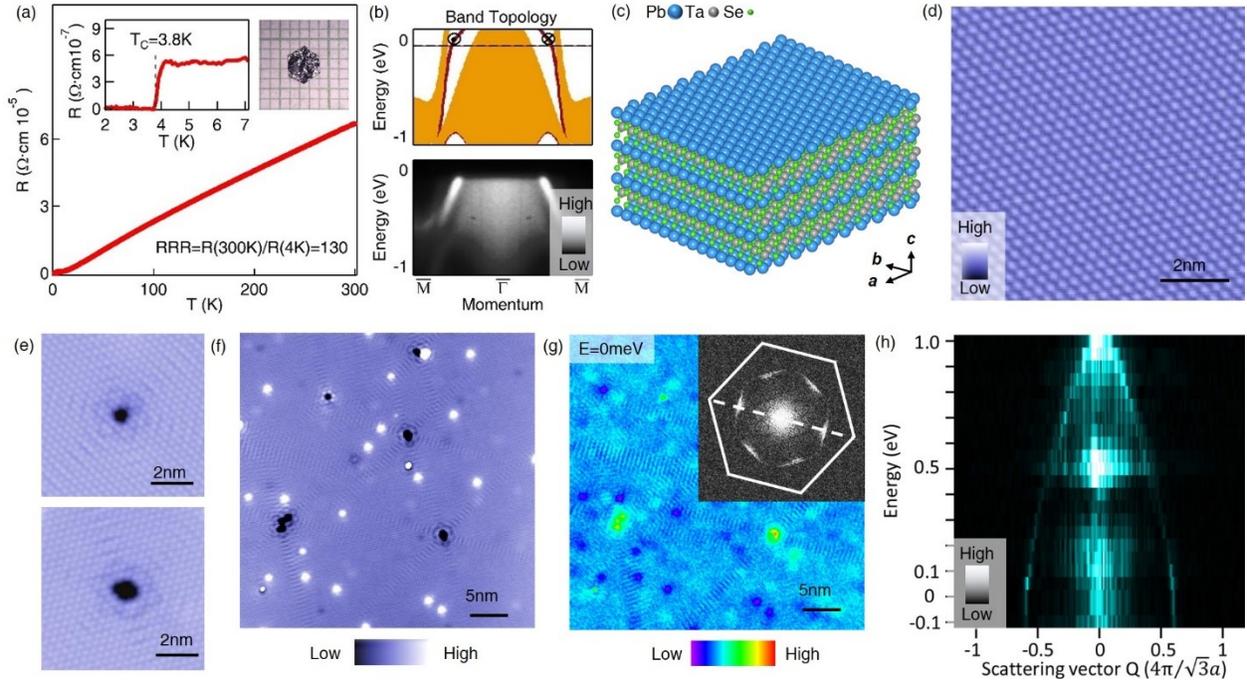

Figure 1. (a) Transport measurement of PbTaSe$_2$ showing a large residual-resistance ratio of 130, indicating the high quality of the sample. Inset left: enlarged resistivity curve near the superconducting transition temperature. Inset right: the optical image of single crystal used in the experiment. (b) Upper: calculated band structure along a high symmetry direction with bulk states (orange) and surface states (red) and spin orientation of the surface states are marked. Lower: photoemission data with photon energy 120 eV, showing the pronounced surface states. (c) 3D view of the crystal structure of PbTaSe$_2$ with a Pb terminating surface. (d) Topographic image of the Pb surface. (e) Topographic image of Pb surfaces with Pb vacancies, which induce anisotropic standing waves. (f) Topographic image of the Pb surface with native impurities. (g) dI/dV map taken at 0meV for the same area. The inset is the Fourier transform of the map showing arc-like QPI signals. The white solid lines in the inset image connect the six Bragg peaks. (h) QPI dispersion along the dash line direction marked in the inset of (g), showing a Dirac cone near 1eV.

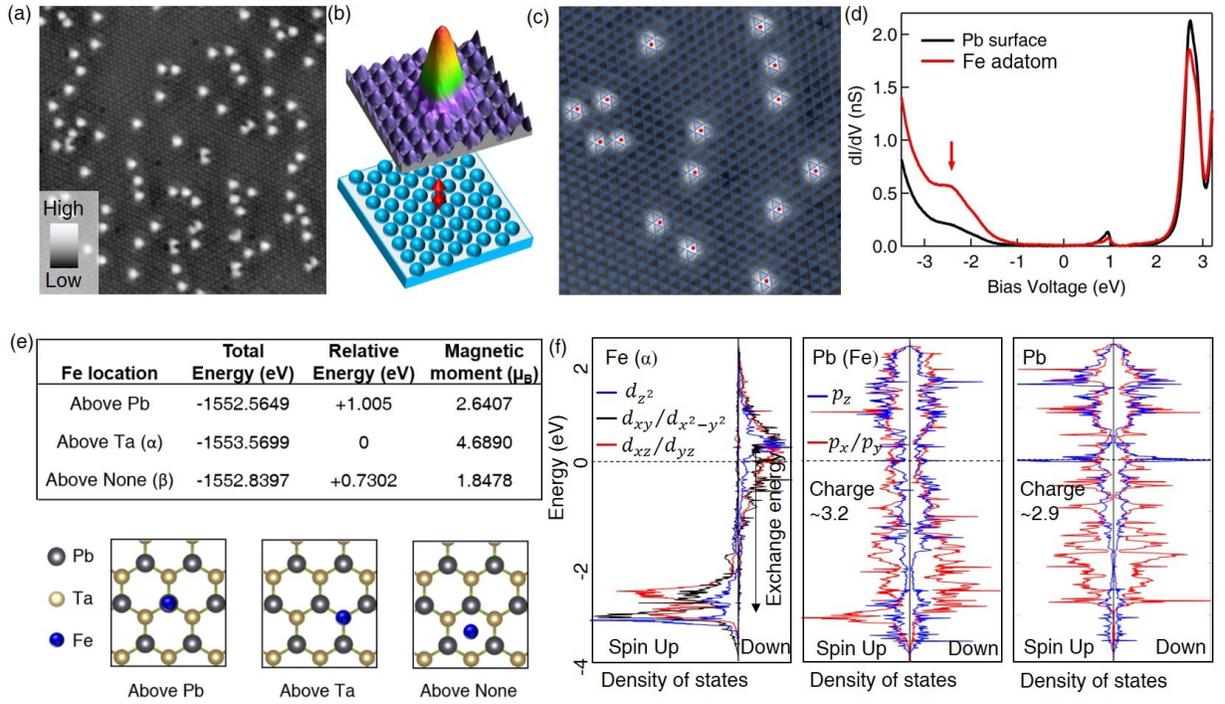

Figure 2. (a) A topographic image showing 0.5% Fe adatoms (randomly scattered bright spots) on the Pb surface. (b) 3D view of an Fe adatom topographic data and its underlying atomic schematic. (c) An overlay of triangle markers on an enlarged region illustrating the two sites of the underlying hexagonal lattice, with unshaded triangles denoting site α and blue shaded triangles denoting site β. The center of each Fe adatom is marked with a red dot. The Fe adatoms all sit at site α. (d) Tunneling spectrums taken at the Fe adatom and away from Fe adatom on Pb surface. Fe adatoms induce additional states around -2.5eV. (e) First-principles calculation of Fe adatom on Pb surface at different absorption sites, showing different energies and magnetic moments. (f) Spin and orbital resolved calculation of the local density of states for Fe adatom and surface Pb atom. We calculated the density of states for the Fe adatom (left panel), the Pb atom nearest to the Fe adatom (middle panel), and Pb atom away from the Fe adatom (right panel). For each panel, we distinguish the spin up (plotted on the left) and spin down states (plotted on the right), and we also distinguish different orbitals for each atom.

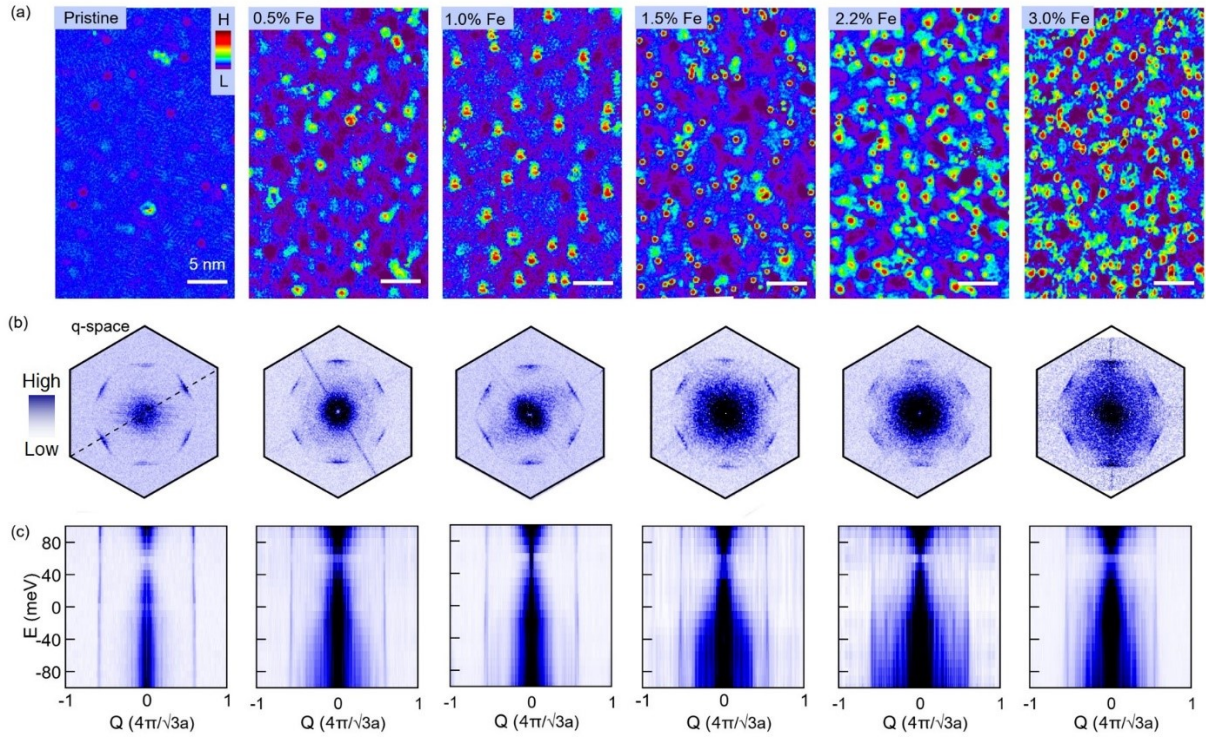

Figure 3. (a) Real space dI/dV maps at zero-energy for different Fe adatom concentrations. (b) Fourier transform of dI/dV maps. The scattering vectors broaden as IFI concentration increases; however, no clear new scattering channels are introduced, and the six-fold arc pattern persists, indicating the robustness of the topological surface state. (c) QPI dispersions taken along the direction marked by the dash lines in (b).